\documentclass[aps,prl,twocolumn,superscriptaddress,nobibnotes]{revtex4-1}
\usepackage[utf8]{inputenc}

\usepackage{color}
\usepackage{graphicx}
\usepackage{amsmath}
\usepackage{amssymb}
\usepackage{bm}
\usepackage{acronym}
\usepackage{ifthen}
\usepackage{blindtext}
\usepackage[normalem]{ulem}
\usepackage{hyperref}
\usepackage[capitalise]{cleveref}
\usepackage{etoolbox}
\usepackage{xspace}
\usepackage{textcomp}
\usepackage{multirow}
\usepackage{lineno}
\usepackage{tabularx}
\usepackage{orcidlink}
\def\gw{GW\xspace}
\def\gwh{GW\xspace}

\def\gws{GWs\xspace}

\def\dm{DM\xspace}
\def\dmh{DM\xspace}
\def\pbh{PBH\xspace}
\def\pbhs{PBHs\xspace}

\def\fh{\emph{frequency-Hough}\xspace}
\def\lvk{LIGO, Virgo and KAGRA\xspace}

\def\mf{matched filtering\xspace}
\def\psd{PSD\xspace}
\def\nn{\nonumber}

\def\cbc{compact binary coalescence\xspace}

\def\gfh{\texttt{GFH}\xspace}
\def\gfhvtwo{\texttt{GFH-v2}\xspace}
\def\BinGFH{\texttt{BinaryGFH}-v2\xspace}

\def\supp{Supplemental Material\xspace}

\def\pmap{\emph{peakmap}\xspace}

\def\erfc{\mathrm{erfc}}

\newcommand{\bea}{\begin{eqnarray}}
\newcommand{\eea}{\end{eqnarray}}
\newcommand{\be}{\begin{equation}}
\newcommand{\ee}{\end{equation}}
\newcommand{\thetathr}{\theta_\text{thr}}

\newcommand{\avgVT}{\ensuremath{\left\langle VT \right\rangle}}

\newcommand{\TFFT}{T_\text{FFT}}
\newcommand{\fgw}{f_\text{GW}}
\newcommand{\fdotgw}{\dot{f}_{\rm GW}}

\newcommand{\Tobs}{T_\text{obs}}
\newcommand{\Tpm}{T_\text{PM}}
\newcommand{\fpbh}{f_\text{PBH}}

\newcommand{\ftilde}{\tilde{f}}

\newcommand{\fsup}{f_\text{sup}}

\newcommand{\Mc}{\mathcal{M}}

\newcommand{\fmin}{f_\text{min}}
\newcommand{\fmax}{f_\text{max}}
\newcommand{\ssm}{sub-solar mass\xspace}
\newcommand{\msun}{\ensuremath{M_\odot}\xspace}
\newcommand{\crthr}{CR_\text{thr}}

\newcommand{\mpbh}{M_\text{PBH}}
\newcommand{\dmax}{d_{\text{max}}}

\newtoggle{fullauthorlist}
\toggletrue{fullauthorlist}
\newtoggle{endauthorlist}
\toggletrue{endauthorlist}

\graphicspath{{./figures/}}

\begin{document}
% \linenumbers
\title{Bridging the chirp-mass gap: searching for gravitational waves from inspiraling subsolar-mass ultra-compact binaries using LIGO O4ab data}
\author{Andrew L. Miller\,\orcidlink{0000-0002-4890-7627}}
\email{andrew.miller.ligo@ucas.ac.cn}
\affiliation{International Centre for Theoretical Physics Asia-Pacific (ICTP-AP), University of Chinese Academy of Sciences (UCAS), Beijing 100190, China.}
\affiliation{Taiji Laboratory for Gravitational Wave Universe, University of Chinese Academy of Sciences, 100049 Beijing, China}
\date{\today}
% \date{\today}

% \label{firstpage}
% \pagerange{\pageref{firstpage}--\pageref{lastpage}}

% Abstract of the paper
\begin{abstract}
Gravitational-wave searches for subsolar-mass compact binaries have left a gap between the planetary-mass regime accessible to long-duration searches and the higher masses probed by conventional compact-binary analyses. We close this gap by searching for the inspiral of ultra-compact binaries with chirp masses $10^{-2}$--$10^{-1}\,M_\odot$ in LIGO O4ab data. Using the statistically robust \texttt{BinaryGFH-v2} method, a time-frequency technique  to search for rapidly evolving inspirals lasting minutes to hours, we cover this previously challenging parameter space with a computational cost of only $\sim 350$ CPU-hours, and find no significant candidates. We thus set 90\% confidence upper limits on the binary merger-rate density, providing the first gravitational-wave constraints across this chirp-mass range. Assuming that primordial-black-hole binaries formed in the early Universe with no binary suppression ($f_{\rm sup}=1$), these results provide the first gravitational-wave constraints on the primordial-black-hole dark-matter fraction in this mass range, yielding upper limits with $f_{\rm PBH}<1$ for equal-mass binaries with component masses $10^{-2}$--$10^{-1}\,M_\odot$ and asymmetric binaries with secondary masses $m_2=3\times10^{-4}$--$10^{-2}\,M_\odot$ for a primary mass $m_1=2.5\,M_\odot$.
\end{abstract}

\maketitle

\section{Introduction}\label{sec:intro}

The detection of low-spinning black holes by \lvk
\cite{aasi2015advanced,acernese2014advanced,KAGRA:2020tym,Abbott:2016blz,TheLIGOScientific:2016pea,Abbott:2016nmj,Abbott:2017vtc,Abbott:2017oio,Abbott:2017gyy,LIGOScientific:2018mvr,LIGOScientific:2020stg,Abbott:2020uma,Abbott:2020khf,Abbott:2020tfl,Abbott:2020mjq}
has renewed interest in primordial black holes (\pbhs) as dark-matter (\dmh) candidates
\cite{Bird:2016dcv,Clesse:2016vqa,Sasaki:2016jop}.
Depending on their formation mechanism in the early Universe
\cite{Carr:2019kxo,Byrnes:2018clq,Jedamzik:2020ypm,Jedamzik:2020omx,DeLuca:2020agl,Escriva:2022duf},
\pbhs could span a broad range of masses and constitute some fraction $\fpbh$, or potentially all, of \dm
\cite{Sasaki:2016jop,Ali-Haimoud:2017rtz,Hall:2020daa,DeLuca:2020jug,Bird:2016dcv,Clesse:2016vqa,Carr:2019kxo}.
However, determining the origin of black holes above one solar mass is difficult because both astrophysical and primordial formation channels can produce such objects
\cite{Clesse:2016vqa,Clesse:2020ghq,Takhistov:2020vxs}.
In contrast, \ssm black holes cannot be explained by stellar evolution processes; therefore, their detection would provide compelling evidence for a primordial origin
\cite{LISACosmologyWorkingGroup:2023njw,Yamamoto:2023tsr}.
Gravitational-wave (\gw) searches for inspiraling \ssm compact binaries thus provide a direct way to test the \pbh hypothesis
\cite{LIGOScientific:2019kan,LIGOScientific:2021job,Phukon:2021cus,Nitz:2021vqh,Nitz:2021mzz,LIGOScientific:2022hai,Nitz:2022ltl}. 

Such searches complement other probes of \pbhs. Indeed, independent constraints on \pbhs have been obtained from microlensing surveys such as HSC \cite{Croon:2020ouk}, OGLE \cite{Niikura:2019kqi}, and EROS \cite{EROS-2:2006ryy}. However, such constraints can depend on the assumed spatial distribution of \pbhs; in particular, clustering can weaken microlensing limits \cite{Garcia-Bellido:2017xvr,Calcino:2018mwh,Belotsky:2018wph,Carr:2019kxo,Trashorras:2020mwn,DeLuca:2020jug}. \gwh searches therefore provide an independent probe of the \pbh population under different formation modeling assumptions.

\gwh searches for \pbh binaries with ground-based detectors tend to divide into two regimes. Matched-filter searches probe binaries approaching the solar-mass scale, but become computationally challenging at chirp masses below $10^{-1}\msun$ because the signals can remain in band for hours to days, requiring prohibitively large template banks \cite{Usman:2015kfa,Sachdev:2019vvd, Kacanja:2026byy,LIGOScientific:2026wxz}. On the other hand, semicoherent time-frequency methods -- methods that break the data into chunks of length $\TFFT$ and combine the power in each fast Fourier transform (FFT) without the phase information -- avoid excessive computational costs and have enabled searches for planetary-mass binaries at lower chirp masses $\Mc$ \cite{Miller:2020kmv,Miller:2021knj,Andres-Carcasona:2023zny,Alestas:2024ubs,Andres-Carcasona:2024jvz,Miller:2024jpo,Miller:2024rca,Rodriguez:2026oyi,Miller:2024fpo,LIGOScientificCollaborationtheVirgoCollaboration:2025cwh,Rodriguez:2026oyi,Wang:2025fmh,Wang:2025lpj,Lu:2026irg}, up to $10^{-2}\msun$. The intermediate range $\Mc=10^{-2}$--$10^{-1}\msun$, however, lies between the regions efficiently covered by these approaches. This gap arises from the qualitatively different signal durations in this
intermediate regime. At lower chirp masses, binaries evolve sufficiently
slowly that long-duration time-frequency techniques can efficiently track
their evolution, whereas at higher masses the shorter signals are well
suited to coherent matched-filter searches. Binaries in the $\Mc=10^{-2}$--$10^{-1}\msun$ intermediate
regime considered here can instead remain in-band for minutes to hours
while evolving by tens of hertz, making them rapidly evolving for
long-duration searches but comparatively expensive to cover with matched filtering. This regime thus provides a
natural target for a semicoherent method that can follow strongly
chirping signals \cite{Miller:2025ote}.

In this \emph{letter}, we search for \gwh signals from inspiraling systems within this chirp-mass range using data from the O4a and O4b observing runs of LIGO \cite{2015CQGra..32g4001L} with \BinGFH \cite{Miller:2025ote}, the statistically robust version of the Generalized frequency-Hough (\gfh) method \cite{Miller:2018rbg,Miller:2024jpo} and its second version \gfhvtwo \cite{SajithMenon:2025vpl}. 
% to search for long-transient inspiral signals. 
Unlike earlier versions of the \gfh, \BinGFH combines the computational efficiency of a semicoherent search with a detection statistic that approximately follows a standard normal distribution in Gaussian noise and in the bulk of detector noise, enabling a statistically rigorous search of detector data to be performed over a wide parameter space that is computationally challenging to cover with matched filtering. 

We do not find any surviving \gwh candidates and place merger-rate upper limits that connect the lower-mass semicoherent searches to higher-mass matched-filtering searches. Then, we interpret these limits as constraints on \pbh binary formation. Assuming no binary suppression, $\fsup=1$, our merger-rate limits, for the first time, constrain $\fpbh<1$ in this mass range for both equal-mass and asymmetric mass-ratio systems in this chirp-mass range. Constraints on these systems can still be obtained for smaller values of $\fsup$, especially for asymmetric mass-ratio systems.

\section{Search}\label{sec:search}

For a quasi-circular compact binary far from merger with negligible spins, the leading-order \gwh frequency evolution is \cite{maggiore2008gravitational}
\begin{equation}
    \fdotgw=\frac{96}{5}\pi^{8/3}\left(\frac{G\Mc}{c^3}\right)^{5/3}\fgw^{11/3}\equiv k\fgw^{11/3},
    \label{eqn:fdot_chirp}
\end{equation}
where $\Mc\equiv(m_1m_2)^{3/5}/(m_1+m_2)^{1/5}$ is the chirp mass, $m_1,m_2$ are the component masses, $k\propto\Mc^{5/3}$ is a proportionality constant, $c$ is the speed of light and $G$ is Newton's gravitational constant.
Integrating \cref{eqn:fdot_chirp}, we obtain the frequency evolution $\fgw(t)$:
\begin{equation}
\fgw(t)=f_0\left[1-\frac{8}{3}kf_0^{8/3}(t-t_0)\right]^{-\frac{3}{8}}~,
\label{eqn:powlaws}
\end{equation}
where $t_0$ is a reference time at the \gwh frequency $f_0$ and $t$ is the time at $\fgw$. 
The amplitude $h_0(t)$ from a source a distance $d$ away  evolves as \cite{maggiore2008gravitational}:

\begin{equation}
h_0(t)=\frac{4}{d}\left(\frac{G \Mc}{c^2}\right)^{5/3}\left(\frac{\pi \fgw(t)}{c}\right)^{2/3}.
\label{eqn:h0}
\end{equation}

Across the mass range considered here, the resulting signals evolve through the detector band on timescales of minutes to hours. We therefore use a semicoherent search rather than coherent matched-filtering. As discussed in \cite{LIGOScientificCollaborationtheVirgoCollaboration:2025cwh}, higher-order post-Newtonian corrections do not alter the waveform beyond one frequency bin, and thus do not need to be considered in our time-frequency analysis.

We analyze LIGO Hanford (H1) and Livingston (L1) data from O4a, spanning 24 May 2023 to 16 January 2024, and O4b, lasting from 10 April 2024 to 28 January 2025. The input data are short fast Fourier transform databases (SFDBs) \cite{Astone:2005fj,Astone:2025sfdb}, which are data products derived by fast Fourier transforming the strain time series in 50\% interlaced chunks of 1024 s. Because the required coherence times for this search are 4 s or less, we inverse Fourier transform the SFDBs and then re-Fourier transform with the desired $\TFFT$. Details regarding calibration, data quality, and data preparation are given in the \supp.

The \BinGFH search operates on \emph{peakmaps}: binary time-frequency representations of the data in which frequency bins that are local maxima and exceed a threshold in equalized power, called ``peaks'', are selected. For the power-law inspiral evolution given in \cref{eqn:fdot_chirp}, the method maps peaks from the detector $t$--$\fgw$ plane into lines in the $f_0$--$\Mc$ plane of the source by summing the number of peaks along tracks corresponding to particular chirp masses and coalescence times \cite{Miller:2018rbg,Miller:2020kmv,Miller:2024jpo,SajithMenon:2025vpl,Pierini:2025gfhv2,miller_2025_17984228}. \BinGFH uses nonuniform grids in frequency and chirp mass and a robust per-pixel critical-ratio statistic ($CR$) designed to be approximately standard normal in Gaussian noise and in the bulk of real detector data. This statistic provides a calibrated measure of significance to select and follow up candidates.

We analyze the $71-169$ Hz frequency range for systems with chirp masses between $\Mc=10^{-2}$--$10^{-1}\msun$. This frequency range comes from maximizing the expected sensitivity to inspiraling systems (see the \supp). In particular, the chirp-mass range is divided into five configurations with $\TFFT=[4,3,2,1,0.5]$ s, chosen so that the intrinsic frequency drift is confined to one frequency bin for the largest chirp mass to which each configuration is sensitive. Knowing the chirp mass and the frequency range, we can calculate the duration of the peakmap in each configuration as $\Tpm=[9226,5190,2307,577,314]$ s. Details on the five configurations are given in the \supp.

Candidates, defined to have parameters $f_0,\Mc,t_0,CR$, are selected independently in H1 and L1 and are required to
meet the following conditions: they must (1) be within three bins of each
other in the chirp-mass/frequency parameter space and (2) have an average
critical ratio $\overline{\rm CR}$ that exceeds the configuration-dependent
threshold $\crthr\sim6$. The numerical value of $\crthr$ corresponds to a
single-detector 1\% false-alarm probability in Gaussian noise, including the
trials factor, and is therefore conservatively applied to the average
H1--L1 $CR$ (see \supp for specific $\crthr$). Applying the first condition leads to $158870$ ($139260$) coincident candidates in O4a (O4b); applying the second one results in zero (one) candidate (see \cref{tab:o4b_followup_candidates}).

In the \supp, we have shown that for each configuration, the $CR$ is approximately normally distributed with a mean of zero and a standard deviation of one, but the tails are non-Gaussian for $CR>5$. Thus, we choose to additionally consider sub-threshold candidates down to an average $\overline{CR}=5$ as significant, resulting in two sub-threshold candidates in O4b (see \cref{tab:o4b_followup_candidates}).

% In both searches, the majority of candidates that meet the aforementioned conditions arise from the $\TFFT=0.5$ s configuration, whose $CR$ values have the largest deviation from a standard normal distribution (see the \supp). This deviation arises because the auto-regressive \psd estimation \cite{Astone:2005fj} relies on a coarse frequency-domain sampling (the frequency resolution is every 2 Hz), which causes fluctuations in the estimate and broadens the distribution of the $CR$. In each of the other configurations, however, the $CR$ follows a normal distribution up to a $CR$ of five, which we validated in both Gaussian \cite{Miller:2025ote} and real noise (see the \supp) for a large number of trials. 

% We account for this broadening in the $\TFFT=0.5$ s configuration, and the non-Gaussian tails of the other configurations' distributions, by following up all sub-threshold candidates with $\overline{CR}\geq5$, making our candidate selection conservative. 
% This is because the measured mean and standard deviation of the $CR$
% distribution for this configuration would imply a higher $\crthr$ than
% the one adopted in our analysis. While this ensures a conservative selection of candidates, it does not recover
% the sensitivity loss caused by the broadened $CR$ distribution: the
% $\TFFT=0.5$~s configuration is therefore less sensitive than expected
% in Gaussian noise. Despite that, the theoretical sensitivity prediction
% agrees with empirical injection studies when the same \psd estimated
% from real data is used in both. See the \supp for more details. 

\begin{table*}[t]
\centering
\begin{tabular}{cccccccccc}
\hline\hline
$T_{\rm PM}$ & $T_{\rm FFT}$ & $t_0$ & $f_0^{\rm H1}$ & $f_0^{\rm L1}$ & $\mathcal{M}^{\rm H1}$ & $\mathcal{M}^{\rm L1}$ & $CR_{\rm H1}$ & $CR_{\rm L1}$ & $\overline{CR}$ \\
(s) & (s) & (GPS s) & \multicolumn{2}{c}{(Hz)} & \multicolumn{2}{c}{$(10^{-2}M_\odot)$} &  &  &  \\
\hline
577 & 1.0 & 1404107934 & 82.00 & 83.00 & 3.020 & 3.020 & 8.92 & 2.53 & 5.73 \\
577 & 1.0 & 1404107934 & 80.00 & 83.00 & 3.020 & 3.020 & 12.26 & 2.53 & 7.39 \\
314 & 0.5 & 1398669161 & 73.00 & 73.00 & 8.721 & 8.721 & 2.56 & 9.26 & 5.91 \\
\hline\hline
\end{tabular}
\caption{Parameters of the three O4b candidates selected for follow-up. The recovered frequencies $f_0^{\rm H1}$ and $f_0^{\rm L1}$ are defined at the common reference time $t_0$ halfway through $\Tpm$. Two candidates likely share the same origin and are both in coincidence with the same candidate in L1. Each candidate has a time--frequency track that intersects frequency bins containing known instrumental lines, combs, or comb teeth \cite{Goetz:2026O4Lines}. In particular, the tracks overlap known calibration lines near $77.7$ Hz in H1 and $78.2$ Hz in L1, numerous teeth of $\sim 2$-Hz combs, and other narrow spectral features in H1. These features can contribute excess power to the \pmap frequency bins. Regardless of these associations with known noise features, we follow up all candidates. }
\label{tab:o4b_followup_candidates}
\end{table*}

We assess whether these candidates are real in the follow-up stage of the search. We demodulate the data using each candidate's expected inspiral phase evolution, accounting for uncertainties in the chirp mass and coalescence time, and double $\TFFT$. A perfectly demodulated signal would appear monochromatic; however, discretization and parameter uncertainties can leave residual frequency drifts. We therefore reanalyze the demodulated data using the independent \fh method \cite{Astone:2014esa}, which searches for lines of different slopes in the time-frequency plane and can thus recover residual spin-downs or spin-ups \cite{Miller:2018rbg}. We select the loudest 1\% of candidates in each sub-band of the \fh map, require the frequencies and spin-downs recovered independently in H1 and L1 to agree within three bins, and retain candidates with $\overline{CR}>5$, where the CR is computed following \cite{Astone:2014esa}, rather than \cite{Miller:2025ote}. We deliberately set this threshold below that corresponding to a 1\% false-alarm probability in the original search to minimize the risk of rejecting a real signal during follow-up. No candidates survive this procedure; see the \supp for further details.

\section{Merger-rate constraints}\label{sec:results}

With no surviving candidates, we determine the sensitive spacetime volume $\avgVT$ from simulated signals distributed uniformly in volume. For 17 chosen chirp masses within the range $10^{-2}-10^{-1}\msun$, we calculate
\begin{equation}
    \avgVT = V_{\rm inj}\,\Tobs\,p_{\rm det},
    \label{eq:VT_injections}
\end{equation}
where $p_{\rm det}=N_{\rm det}/N_{\rm inj}$ is the recovered fraction, $N_{\rm det}$ is the number of recovered signals, $V_{\rm inj}=4\pi \dmax^3/3$, and $\dmax$ is chosen separately for each chirp mass
to balance computational cost against adequately sampling the nonzero-efficiency regime, avoiding unnecessary injections at distances where the detection efficiency is zero. $\dmax$ increases from \(1.22\,\mathrm{Mpc}\) at the low-mass end to \(5.88\,\mathrm{Mpc}\) at the high-mass end of the chirp-mass range.

$\Tobs$ is the effective observing time calculated from the available coincident L1--H1 data (see the \supp for more details on $\Tobs$ and $\dmax$). We require recovered injections to pass the same $\overline{CR}>5$ sensitivity floor used to select candidates for follow-up. The derived $\avgVT$ values for each configuration are shown in the \supp. We note here that for the $\TFFT=0.5$~s configuration, the coarse resolution in frequency impacts the power spectral density (\psd) estimation,
reducing the sensitivity relative to that expected in Gaussian noise.
This loss is captured directly by the $\avgVT$ derived with injections and is
consistent with the theoretical sensitivity predicted using the \psd
estimated from the data for this configuration. Details of the injection distributions, efficiency curves, \psd estimation, and validation of the sensitive volume are given in the \supp.

Assuming Poisson-distributed mergers, the 90\% confidence upper limit can be calculated from $\avgVT$ \cite{LIGOScientificCollaborationtheVirgoCollaboration:2025cwh}
\begin{equation}
    \mathcal{R}_{90\%}=\frac{2.303}{\avgVT}.
    \label{eqn:R90}
\end{equation}

\begin{figure*}[ht]
    \centering
    \includegraphics[width=0.75\textwidth]{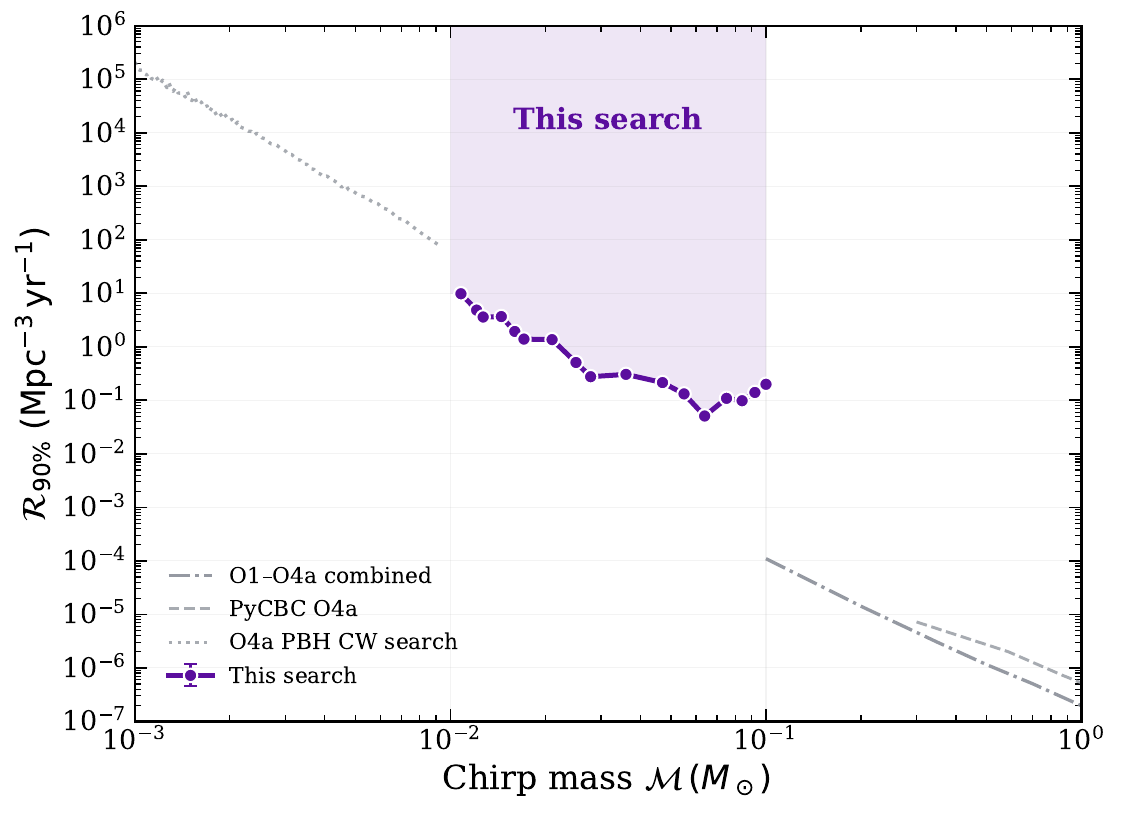}
    \caption{90\% confidence upper limits on the binary merger-rate density as a function of chirp mass. Our search covers the $\Mc=10^{-2}$--$10^{-1}\msun$ range, which lies between previous long-duration, semicoherent searches for planetary-mass binaries \cite{LIGOScientificCollaborationtheVirgoCollaboration:2025cwh,LIGOScientific:2026plm} and higher-mass matched-filtering searches \cite{LIGOScientific:2026wxz,Kacanja:2026byy}. The shaded area shows the parameter space probed in this work. At each chirp mass, $10^4$ injections were performed for 17 chirp masses between $10^{-2}-10^{-1}\msun$.}
    \label{fig:R90_high_chirp_mass}
\end{figure*}

\cref{fig:R90_high_chirp_mass} places our limits on $\mathcal{R}_{90\%}$ in the context of previous \gw searches. Long-duration searches with other semicoherent methods probe chirp masses below $\sim10^{-2}\msun$, whereas matched-filter searches constrain the region above $\sim10^{-1}\msun$. Our analysis fills this mass gap, and connects the \gwh merger-rate constraints across these previously separated search regimes. 

Importantly, accessing this region does not require the computational
cost of a matched-filtering search. The complete
\BinGFH analysis of O4a and O4b requires only $\sim 350$
CPU-hours, thus demonstrating that the intermediate regime can be systematically
probed with semicoherent time-frequency methods. 

\section{Primordial-black-hole constraints}\label{sec:pbh}

Converting merger-rate limits into a constraint on the \pbh \dmh abundance is necessarily model dependent. To do so, we use the early-Universe two-body binary-formation prescriptions of Ref.~\cite{Raidal:2024bmm}, for which the cosmological merger-rate density can be written as
\begin{align}
    \mathcal{R}^{\rm cos}_{\rm prim}
    &\approx 1.6\times10^{-3}\,\mathrm{Mpc}^{-3}\mathrm{yr}^{-1}\,\ftilde^{53/37} \nonumber\\
    &\quad\times\left(\frac{m_1+m_2}{\msun}\right)^{-32/37}
    \left[\frac{m_1m_2}{(m_1+m_2)^2}\right]^{-34/37},
    \label{eq:cosmomerg}
\end{align}
where
\begin{equation}
    \ftilde \equiv \fpbh
    \left[\fsup f(m_1)\Delta m_1 f(m_2)\Delta m_2\right]^{37/53}.
    \label{eqn:ftilde}
\end{equation}
Here, $f(m)$ is the normalized \pbh mass probability density, $\Delta m$ is the width of the mass bin, and $\fsup$ accounts for processes that suppress the primordial-binary merger rate \cite{Raidal:2018bbj}. Expressing the result in terms of the effective parameter $\ftilde$ collapses the uncertainties associated with the mass function and suppression factor into one number.

As in \cite{LIGOScientificCollaborationtheVirgoCollaboration:2025cwh}, we convert the cosmological merger rate to a local rate using the Galactic \dmh density profile, which results in an enhancement due to the local \dmh energy density. Additionally, in this search, our reach extends beyond tens of kiloparsecs to approximately a few Mpc, such that the M31 \dmh halo $\sim 770$ kpc away from us can also enhance the local merger rate \cite{10.1093/mnras/stae025}. However, for both \dmh profiles, we must account for how this enhancement will fall off with distance from the center of each galaxy. See the \supp for more details. 

\begin{figure}[ht]
    \centering
    \includegraphics[width=\columnwidth]{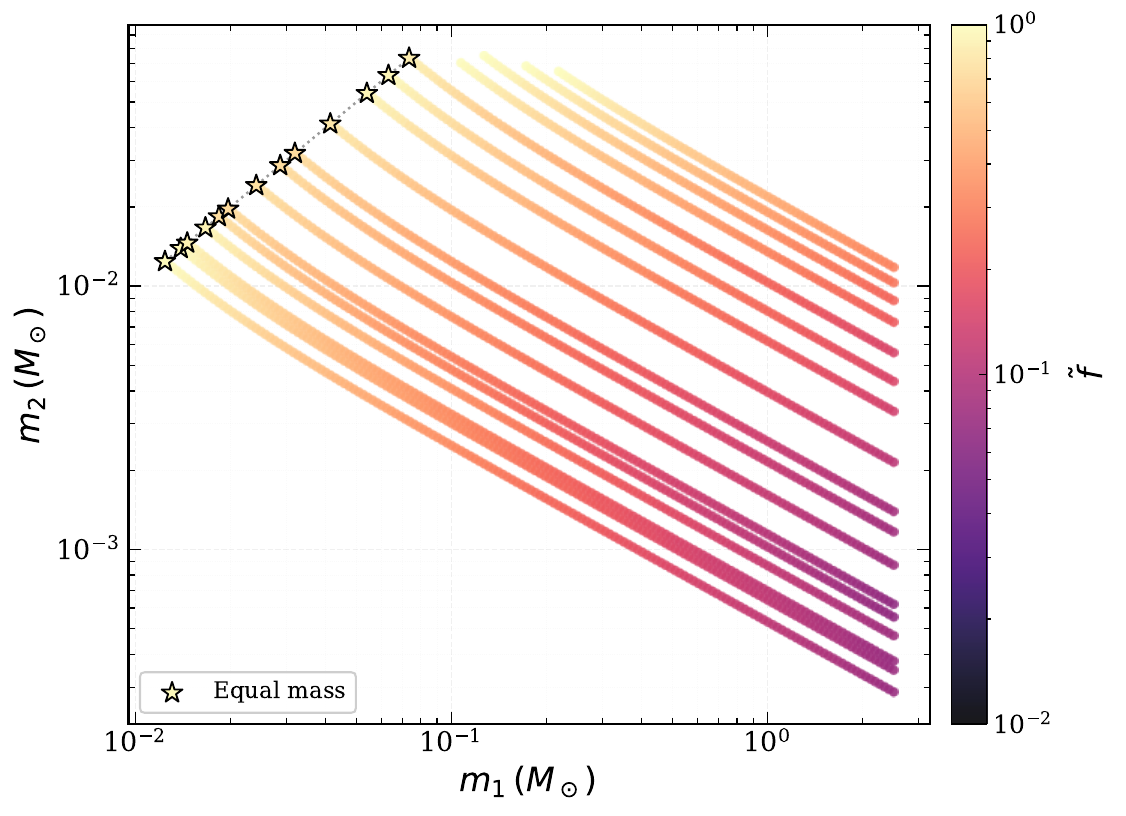}
\caption{
90\% confidence upper limits on the effective
parameter $\ftilde$ inferred from the O4ab merger-rate constraints.
$\ftilde$ combines the \pbh \dmh fraction, mass function, and
suppression of the primordial-binary merger rate (see \cref{eqn:ftilde});
smaller values therefore correspond to stronger constraints on \pbh
binary formation. Curves show asymmetric systems as a function of
component masses $m_1$ and $m_2$, while stars denote equal-mass systems.
These limits do not fix a particular \pbh mass function or binary
suppression factor and can be mapped onto $\fpbh$ for different choices
of these quantities.
    }
    \label{fig:ftilde_constraints}
\end{figure}

By specializing \cref{eq:cosmomerg} to equal- and asymmetric mass-ratio systems, we can convert $\mathcal{R}_{90\%}$ to upper limits on $\ftilde$, as shown in \cref{fig:ftilde_constraints}. These upper limits are derived without assuming a \pbh mass function or a binary suppression factor in the 
early Universe. 

$\ftilde$ can be used to constrain $\fpbh$
for different assumptions about the \pbh mass function and binary
suppression without repeating this analysis, permitting our results to be used by the \pbh community. We thus present
$\ftilde$ as the primary constraint in this work,
and choose $f(m)$ and $\fsup$ to illustrate how $\fpbh$ can be derived. 
% of the mass function and $\fsup$ to
% illustrate the corresponding implications for the \pbh \dmh
% fraction.

% It thus provides a basis from which the community can constrain their own \pbh hypotheses.

\begin{figure}[ht]
    \centering
\includegraphics[width=\columnwidth]{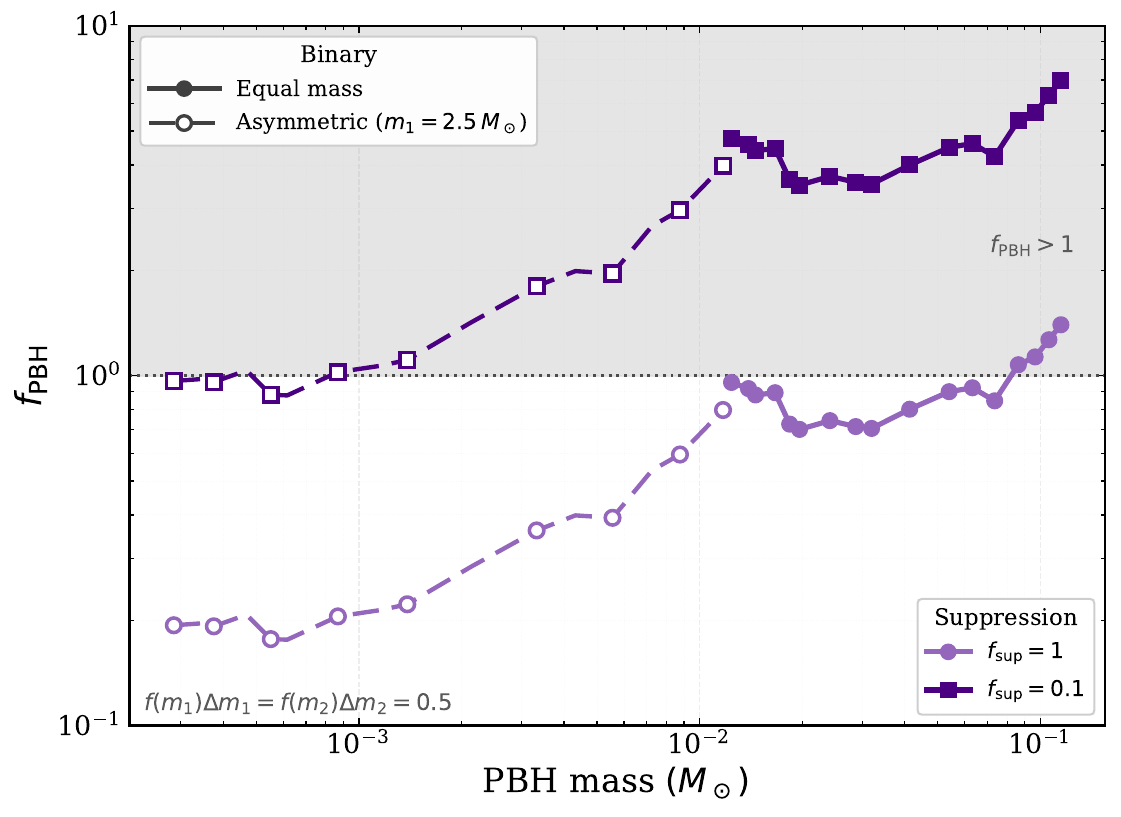}
\caption{Constraints on the \pbh \dmh fraction $\fpbh$ obtained from the O4ab merger-rate limits. Circles correspond to constraints assuming no binary suppression, while  squares indicate constraints with a suppression factor of $\fsup=0.1$. Colored markers indicate constraints on equal-mass systems, while open markers refer to constraints on asymmetric mass-ratio systems that assume a primary mass of $m_1=2.5\msun$ motivated by the QCD phase transition. To obtain these constraints from $\ftilde$, we assume that half of black holes are at $m_1$, while the other half are at the plotted $m_2$ values. The shaded region corresponds to the regime in which $\fpbh>1$.
    }
    \label{fig:fpbh_constraint}
\end{figure}

To compare with other \pbh abundance limits, we assume that the \pbh mass distributions are $f(m_1)\Delta m_1=f(m_2)\Delta m_2=f(\mpbh)\Delta\mpbh=1$ for the equal-mass case and  $f(m_1)\Delta m_1=f(m_2)\Delta m_2=0.5$ for asymmetric mass-ratio systems assuming $m_1=2.5\msun$, motivated by the \pbh masses that would be enhanced during the QCD transition \cite{Byrnes:2018clq,Carr:2019kxo,Clesse:2020ghq}. For zero ($\fsup=1$) and some ($\fsup=0.1)$ merger-rate suppression, we show in \cref{fig:fpbh_constraint} constraints on $\fpbh$ in the mass range probed by our search. Our results provide the first \gwh constraints in the chirp-mass range of $10^{-2}-10^{-1}\msun$ for $\fpbh<1$ in both of the aforementioned suppression cases for asymmetric mass-ratio systems, and showcase the first constraints on equal-mass systems in the mass range with $\fpbh<1$ for $\fsup=1$. 

\section{Conclusions}\label{sec:concl}

We searched LIGO O4a and O4b data for long-transient inspiral signals from ultra-compact binaries with chirp masses $\Mc=10^{-2}$--$10^{-1}\msun$. This interval lies between the parameter spaces previously covered by long-duration planetary-mass searches and conventional compact-binary searches. Using the statistically robust \BinGFH method, we find no surviving \gw candidates and set 90\% confidence upper limits on the merger-rate density across this previously separated region.

The search demonstrates that our method, using only $350$ CPU-hours, can efficiently constrain long-transient, subsolar-mass binary signals that are computationally costly to probe with matched filtering. This cost is approximately $2-3$ orders of magnitude lower than that employed by the \gfhvtwo in \cite{LIGOScientificCollaborationtheVirgoCollaboration:2025cwh} and $4-5$ orders of magnitude lower than matched-filtering searches in \cite{LIGOScientific:2026wxz}. In particular, the new CR statistic approximately follows a standard normal distribution in the bulk of detector noise, providing statistical robustness and avoiding the uncalibrated CR values in the tails encountered in \cite{LIGOScientificCollaborationtheVirgoCollaboration:2025cwh}.

Physically, the resulting rate-density limits bridge the gap between existing \gwh constraints below and above  the $10^{-2}$--$10^{-1}\msun$ chirp-mass range. Moreover, extrapolating the sensitivity of the previous O4a PBH continuous-wave search into this chirp-mass range using its expected power-law scaling yields merger-rate limits roughly an order of magnitude weaker than those obtained here. This demonstrates a significant improvement in sensitivity using \BinGFH relative to that of \gfhvtwo. While \mf in this mass range would achieve greater sensitivity, its computational cost makes such a search computationally challenging, underscoring the importance of dedicated methods for probing inspiraling binaries in this regime.

By assuming that \pbhs formed binaries in the early Universe, we can constrain the effective abundance $\ftilde$ for asymmetric binaries. For $\fsup=1$, our results also provide the first \gwh constraints on $\fpbh$ in this mass range for both equal- and asymmetric mass-ratio binaries. However, the degree of merger-rate suppression is uncertain, and thus our constraints, as well as other limits that arise from the non-detection of \gws in the \ssm regime, must be interpreted carefully.

While the constraints that we obtain do not surpass those from microlensing experiments, they are derived under a different set of astrophysical assumptions. In particular, we probe \ssm objects that could form in binaries, while microlensing experiments are sensitive to isolated ones. Moreover,
each method or experiment that is used to set constraints on $\fpbh$ has its own systematics and assumptions, and it may happen that microlensing constraints weaken
due to a better understanding of the galactic rotation
curves \cite{Garcia-Bellido:2024yaz}. This underscores the need to have simultaneous probes of \pbhs.

% As future ground- and space-based \gwh \ifos extend sensitivity to lower frequencies, binaries will be observed earlier in their evolution, where effects such as orbital eccentricity \cite{Ramos-Buades:2021adz,Liu:2023ldr,Gamba:2024cvy,Nagar:2024dzj,Gamboa:2024hli,Bhaumik:2024cec,Morras:2025xfu,Morras:2025nlp,Planas:2025feq,Huez:2025npe} and environmental interactions \cite{Bertone:2019irm,Kavanagh:2020cfn,Cole:2022yzw,Kavanagh:2024lgq,Karydas:2024fcn,Dyson:2025dlj,Mitra:2025tag} may become increasingly important. However, because the \BinGFH tracks power in the time-frequency plane rather than requiring phase coherence within a template over the full signal duration, it may be less sensitive to waveform mismodeling from such effects. This flexibility, together with its low computational cost, makes semicoherent searches particularly promising for probing a broad range of compact binary coalescences. 

\section*{Data Availability Statement}
The data and code that support these findings can be found in a Zenodo repository \cite{miller_2026_22729157}.

\section*{Acknowledgments}
This material is based upon work supported by NSF's LIGO Laboratory which is a major facility fully funded by the National Science Foundation.

This work is supported by the National Natural Science Foundation of China (NSFC) under Grant Nos. 12347103, 12547104, and W2611007.

We would like to thank the Rome Virgo group for the tools necessary to perform these studies, such as the development of the original \fh transform and the development of the short FFT databases. Additionally we would like to thank Pia Astone, Lorenzo Silvestri and Stefano Dal Pra for support with the use of SFDBs and the CNAF computing center.

This research has made use of data, software and/or web tools obtained from the Gravitational Wave Open Science Center (https://www.gw-openscience.org/ ), a service of LIGO Laboratory, the LIGO Scientific Collaboration and the Virgo Collaboration. LIGO Laboratory and Advanced LIGO are funded by the United States National Science Foundation (NSF) as well as the Science and Technology Facilities Council (STFC) of the United Kingdom, the Max-Planck-Society (MPS), and the State of Niedersachsen/Germany for support of the construction of Advanced LIGO and construction and operation of the GEO600 detector. Additional support for Advanced LIGO was provided by the Australian Research Council. Virgo is funded, through the European Gravitational Observatory (EGO), by the French Centre National de Recherche Scientifique (CNRS), the Italian Istituto Nazionale della Fisica Nucleare (INFN) and the Dutch Nikhef, with contributions by institutions from Belgium, Germany, Greece, Hungary, Ireland, Japan, Monaco, Poland, Portugal, Spain.

We also wish to acknowledge the support of the INFN-CNAF computing center for its help with the storage and transfer of the data used in this paper.

We would like to thank all of the essential workers who put their health at risk during the COVID-19 pandemic, without whom we would not have been able to complete this work.

\bibliographystyle{apsrev4-1}
\bibliography{references} 

\clearpage

\onecolumngrid
% \vspace*{-1em}

\begin{center}
    \textbf{\huge Supplemental Material}

    \medskip
    \rule{\textwidth}{1.5pt}
\end{center}

\medskip
\twocolumngrid

\appendix

% ============================================================
\section{Data set and search configuration}
\label{app:data-search}
% ============================================================

We analyze data from the LIGO Hanford (H1) and Livingston (L1)
detectors during O4a and O4b. O4a spans 24 May 2023 15:00:00 UTC
(GPS 1368975618) to 16 January 2024 16:00:00 UTC
(GPS 1389456018), with duty cycles of 67.5\% and 69\% for H1 and
L1, respectively. O4b spans 10 April 2024 15:00:00 UTC
(GPS 1396796418) to 28 January 2025 17:00:00 UTC
(GPS 1422118818), with duty cycles of 48.6\% and 68.1\% for H1 and
L1, respectively. We use coincident science-mode data and the
\textsc{GDS--CALIB\_STRAIN\_CLEAN\_AR} strain channel with
\textsc{CAT1} vetoes \cite{LIGO:2024kkz}. The two LIGO detectors
operated simultaneously for approximately $126.6$~days ($53.4\%$)
during O4a and $112.7$~days ($38.5\%$) during O4b, where the latter
includes periods of three-detector LHO--LLO--Virgo operation
\cite{LIGOScientific:2026gwt}. However, the effective observing time entering the search is configuration dependent because only complete peakmaps of duration $\Tpm$ that are available in both detectors are retained. The resulting values are given in \cref{tab:o4a_search_summary,tab:o4b_search_summary}. The O4 calibration procedure
and uncertainties are described in
Refs.~\cite{Karki:2016pht,Viets:2017yvy,Wade:2025tgt,Glanzer:2026fhy}.

The search input consists of short fast Fourier transform databases
(SFDBs) \cite{Astone:2005fj,Astone:2025sfdb}. These contain
frequency-domain representations of the detector data in
$\sim17$-min intervals and are cleaned of short time-domain
disturbances. Because the coherence times required in this search are much
shorter than 17 minutes, we inverse Fourier transform the SFDB data and construct new
time-frequency representations with the desired $\TFFT$.

We search the frequency band $71$--$169$ Hz and chirp mass range
$\Mc=10^{-2}$--$10^{-1}\msun$. The frequency band is chosen by
maximizing \cite{Alestas:2024ubs}
\begin{equation}
F(\fmin,\fmax)=
\frac{\fmin^{2/3}}{\fmax^{11/24}}
\left[
\int_{\fmin}^{\fmax}
\frac{df}{f^{7/3}S_n(f)}
\right]^{1/2},
\label{eq:supp_frequency_band}
\end{equation}
using a representative O4a LIGO noise power spectral density $S_n(f)$. This equation is independent of both the observing run and the chirp mass; however, there are slight changes in $\fmin,\fmax$ depending on the $S_n(f)$ used. For simplicity, we use [71,169] Hz for both O4a and O4b.

After fixing the frequency, the coherence time is selected primarily based on the chirp mass
\begin{equation}
\TFFT^{\rm opt}
=
8.50~{\rm s}
\left(\frac{\Mc}{10^{-2}\msun}\right)^{-5/6}
\left(\frac{\fmax}{126.8~{\rm Hz}}\right)^{-11/6},
\label{eq:supp_TFFT_opt}
\end{equation}
which leads to five search configurations. The \pmap duration
$\Tpm$ is set by the largest chirp mass covered by each configuration, by inverting \cref{eqn:powlaws}, solving for $(t-t_0)$ and specifying the frequency band:

\begin{equation}
\Tpm
\simeq
314~{\rm s}
\left(
\frac{\Mc}{0.1\,M_\odot}
\right)^{-5/3},
\qquad
f\in[71,169]~{\rm Hz}.
\end{equation}

Tables~\ref{tab:o4a_search_summary} and \ref{tab:o4b_search_summary} give the complete
O4a and O4b configurations and candidate vetoes after thresholding. For each
configuration,  $\Tobs$ is the effective coincident observation
time for which both detectors contain science-mode, non-zero data.
The single-detector trials-factor-corrected threshold $\crthr$ corresponds to a 1\%
false-alarm probability under the nominal standard-normal $CR$
background, accounting for
$N_{\rm pts}=N_{t_0}N_fN_{\Mc}$ searched points, where $N_{t_0}$ is the number of peakmaps analyzed in a given configuration, $N_f$ is the number of frequencies in the Hough-map grid, and $N_{\Mc}$ is the number of chirp masses in the Hough-map grid. This threshold choice is conservative in the sense that the average of two standard normal distributions would result in a reduction of the threshold by $\sqrt{2}$, both increasing the sensitivity of the search (because we keep candidates with even lower $\overline{CR}$) but also increasing the computational cost of the follow-up, potentially significantly, based on the number of candidates present at $CR$s of $\sim 4$ (see the last section for more details).

\begin{table*}[t]
\centering
\begin{tabular}{ccccccc||c||c||c}
\hline\hline
$T_{\rm PM}$ & $T_{\rm FFT}$ & $\mathcal{M}^{\min}$ & $\mathcal{M}^{\max}$ & $N_{\rm pts}$ & $T_{\rm obs}$ & $CR_{\rm thr}$ & Coincident & $\overline{CR} \geq CR_{\rm thr}$ & $5 \leq \overline{CR} < CR_{\rm thr}$ \\
(s) & (s) & \multicolumn{2}{c}{$(10^{-2}M_\odot)$} & $(10^{7})$ & (days) &  &  &  &  \\
\hline
9226 & 4.0 & 1.00 & 1.31 & 2.80 & 91.62 & 6.16 & 1865 & 0 & 0 \\
5190 & 3.0 & 1.31 & 1.85 & 4.06 & 104.70 & 6.22 & 2800 & 0 & 0 \\
2307 & 2.0 & 1.85 & 3.02 & 6.44 & 116.39 & 6.29 & 4255 & 0 & 0 \\
577 & 1.0 & 3.02 & 6.93 & 11.90 & 123.54 & 6.39 & 17466 & 0 & 0 \\
314 & 0.5 & 6.93 & 10.00 & 2.40 & 124.60 & 6.14 & 132484 & 0 & 0 \\
\hline
\textbf{Total} & & & & & & & \textbf{158870} & \textbf{0} & \textbf{0} \\
\hline\hline
\end{tabular}
\caption{How the number of candidates decreases as a function of the cuts that we make in the O4a search. For each configuration, there is a chosen \pmap duration $\Tpm$, coherence time $\TFFT
$, chirp-mass range $\Mc^{\min},\Mc^{\max}$, number of points in the parameter space $N_{\rm pts}$, coincident observation time $\Tobs$, and detection-statistic threshold $\crthr$. ``Coincident'' requires that candidates in L1 and H1 be within 3 bins of each other. $\overline{CR}\geq\crthr$ requires that the average $CR$ in H1 and L1 be greater than the configuration-dependent $\crthr$ given in the table. The final column collects the number of sub-threshold candidates down to $\overline{CR}=5$ in each configuration.}
\label{tab:o4a_search_summary}
\end{table*}

\begin{table*}[t]
\centering
\begin{tabular}{ccccccc||c||c||c}
\hline\hline
$T_{\rm PM}$ & $T_{\rm FFT}$ & $\mathcal{M}^{\min}$ & $\mathcal{M}^{\max}$ & $N_{\rm pts}$ & $T_{\rm obs}$ & $CR_{\rm thr}$ & Coincident & $\overline{CR} \geq CR_{\rm thr}$ & $5 \leq \overline{CR} < CR_{\rm thr}$ \\
(s) & (s) & \multicolumn{2}{c}{$(10^{-2}M_\odot)$} & $(10^{7})$ & (days) &  &  &  &  \\
\hline
9226 & 4.0 & 1.00 & 1.31 & 2.12 & 69.41 & 6.12 & 1380 & 0 & 0 \\
5190 & 3.0 & 1.31 & 1.85 & 3.26 & 84.04 & 6.19 & 2214 & 0 & 0 \\
2307 & 2.0 & 1.85 & 3.02 & 5.35 & 96.77 & 6.26 & 3544 & 0 & 0 \\
577 & 1.0 & 3.02 & 6.93 & 10.20 & 105.89 & 6.36 & 15028 & 1 & 1 \\
314 & 0.5 & 6.93 & 10.00 & 2.06 & 107.16 & 6.11 & 117094 & 0 & 1 \\
\hline
\textbf{Total} & & & & & & & \textbf{139260} & \textbf{1} & \textbf{2} \\
\hline\hline
\end{tabular}
\caption{Same as \cref{tab:o4a_search_summary}, but in O4b.}
\label{tab:o4b_search_summary}
\end{table*}

% ============================================================
\section{Candidate selection and follow-up}
\label{app:final-search}
% ============================================================

The \BinGFH analysis is performed independently on H1 and L1 data.
Each candidate is characterized by a reference frequency, a reference time, chirp mass,
and critical ratio $CR$. The Hough transform is constructed in the
coordinates $x=1/f^{8/3}$ and $k$, where $k$ parametrizes the chirp
mass \cite{Miller:2018rbg}. We define the distance between H1 and L1
candidates as \cite{Astone:2014esa,Miller:2018rbg,Miller:2024jpo}
\begin{equation}
\mathrm{dist}
=
\sqrt{
\left(
\frac{k_{\rm H1}-k_{\rm L1}}{\delta k}
\right)^2
+
\left(
\frac{x_{0,\rm H1}-x_{0,\rm L1}}{\delta x_0}
\right)^2
},
\label{eqn:dbin}
\end{equation}
where $\delta k$ and $\delta x_0$ are the corresponding Hough-map bin
sizes. 
Both of these sizes change as a function of $k$ and $x_0$, respectively, making them both nonuniform.

For each Hough-map bin $(x_0,k)$, we calculate our detection statistic, the critical ratio, as \cite{Miller:2025ote}
\begin{equation}
CR(x_0,k)=
\frac{n(x_0,k)-\mu(x_0,k)}
{\sigma(x_0,k)},
\label{eq:cr}
\end{equation}
where $n(x_0,k)$ is the observed number count and $\mu(x_0,k)$ and
$\sigma(x_0,k)$ are its expected mean and standard deviation in noise.
Unlike previous implementations of the \gfh, these quantities are
evaluated separately for each Hough-map pixel, accounting for the
different numbers of time--frequency bins contributing to different templates, i.e. tracks in the time-frequency plane \cite{Miller:2025ote}.

The probability $p_0$ that a time--frequency bin is selected as a peak
is estimated empirically from the \pmap as the fraction of available
bins that survive the peak-selection procedure. For a Hough bin
receiving contributions from $N_{\rm pairs}(x_0,k)$ time--frequency
bins, the expected mean and variance are \cite{Miller:2025ote}
\begin{align}
\mu(x_0,k) &=
N_{\rm pairs}(x_0,k)\,p_0, \label{eqn:mu}\\
\sigma^2(x_0,k) &=
N_{\rm pairs}(x_0,k)\,p_0(1-p_0)\label{eqn:sig}.
\end{align}
The resulting $CR$ therefore accounts for both variations in the
number of contributing time--frequency bins across the Hough map and
the peak density measured directly from the data. In Gaussian noise and the bulk of real noise, the
$CR$ follows an approximately standard normal
distribution, allowing the statistical significance of candidates to
be quantified \cite{Miller:2025ote}.

We keep candidates separated by no more than three bins. The number of bins to choose is a compromise between retaining sensitivity and minimizing computational cost in the follow-up.
Moreover, for each configuration, the threshold on the critical ratio, $\crthr$, is chosen to
give a 1\% false-alarm probability after accounting for the
look-elsewhere effect \cite{Miller1981}. As discussed in the last section, the measured detector backgrounds have
non-Gaussian positive tails, particularly for the shortest $\TFFT$ we consider.
Consequently, in addition to candidates exceeding $\crthr$, we follow up
all coincident candidates down to an average critical ratio of $\overline{CR}=5$. We show the surviving candidates up until this point in \cref{tab:o4b_followup_candidates}, and report the number of candidates remaining after each veto in \cref{tab:o4a_search_summary,tab:o4b_search_summary}.

In the follow-up phase of the search, each surviving candidate is then demodulated using the phase evolution
predicted by its recovered inspiral parameters. If those parameters
correspond to a real signal, the correction should make the signal monochromatic. We therefore reconstruct the \pmap
with a coherence time twice that used in the original search and
apply the \fh transform to search over any residual
frequency derivative \cite{Astone:2014esa,Miller:2018rbg}. This method maps points in the time-frequency plane of the detector to lines in the frequency-spin-down plane of the source, and can handle residual frequency drifts due to imperfect demodulations \cite{Miller:2018rbg,Miller:2024jpo}. The recovered parameters are used for a final
correction and projection of the \pmap onto frequency. We set a separate threshold $\crthr=5$ on the output of the \fh; however, no candidate satisfies this requirement. 

The main O4a/O4b search required approximately 350 CPU-hours; the
coincidence and follow-up stages were negligible in comparison. The
injection campaigns described below required approximately 4000
CPU-hours in total.

% ============================================================
\section{Validation of the search sensitivity}
\label{app:sensitivity-validation}
% ============================================================

As in \cite{LIGOScientificCollaborationtheVirgoCollaboration:2025cwh}, we validate the measured sensitivity against the analytic expectation
for a \pmap-based search. For each detector, the expected distance
reach at detection probability $\Gamma$ is
\begin{align}
d^\Gamma_{\rm max}
&=
1.41
\left(\frac{G\Mc}{c^2}\right)^{5/3}
\left(\frac{\pi}{c}\right)^{2/3}
\frac{\TFFT}{\Tpm^{1/2}}
\left(\frac{p_0(1-p_0)}{Np_1^2}\right)^{-1/4}
\nonumber\\
&\quad\times
\left(
\sum_x^N
\frac{f_{{\rm GW},x}^{4/3}}
{S_n(f_{{\rm GW},x})}
\right)^{1/2}
\left[
\overline{CR}_{\rm thr}-\sqrt{2}\,\erfc^{-1}(2\Gamma)
\right]^{-1/2},
\label{eqn:dmax}
\end{align}
where $c$ is the speed of light, $\Mc$ is the chirp mass, $\Gamma=0.95$, $N=\Tpm/\TFFT$, $\Tpm$ is the peakmap duration, $\TFFT$ is the coherence time, $\overline{CR}_{\rm thr}=5$ is the sub-threshold detection-statistic value above which all surviving candidates are followed up and vetoed, and $p_0$ is the probability of
selecting a peak above the equalized-power threshold
$\thetathr=2.5$:
\begin{equation}
p_0
=
e^{-\thetathr}
-e^{-2\thetathr}
+\frac{1}{3}e^{-3\thetathr},
\label{eqn:p0}
\end{equation}
with
\begin{align}
p_1
&=
\thetathr
\left(
\frac{1}{2}e^{-\thetathr}
-\frac{1}{2}e^{-2\thetathr}
+\frac{1}{6}e^{-3\thetathr}
\right)
\nonumber\\
&\quad+
\frac{1}{4}e^{-2\thetathr}
-\frac{1}{9}e^{-3\thetathr}.
\label{eqn:p1}
\end{align}
These quantities describe peak selection in the presence of a weak
monochromatic signal \cite{Palomba2025_peakmap}. To compute the theoretical distance reach, we use the \psd $S_n(\fgw)$ obtained directly from the noise. 

\begin{figure}[ht!]
    \centering
    \includegraphics[width=\columnwidth]
    {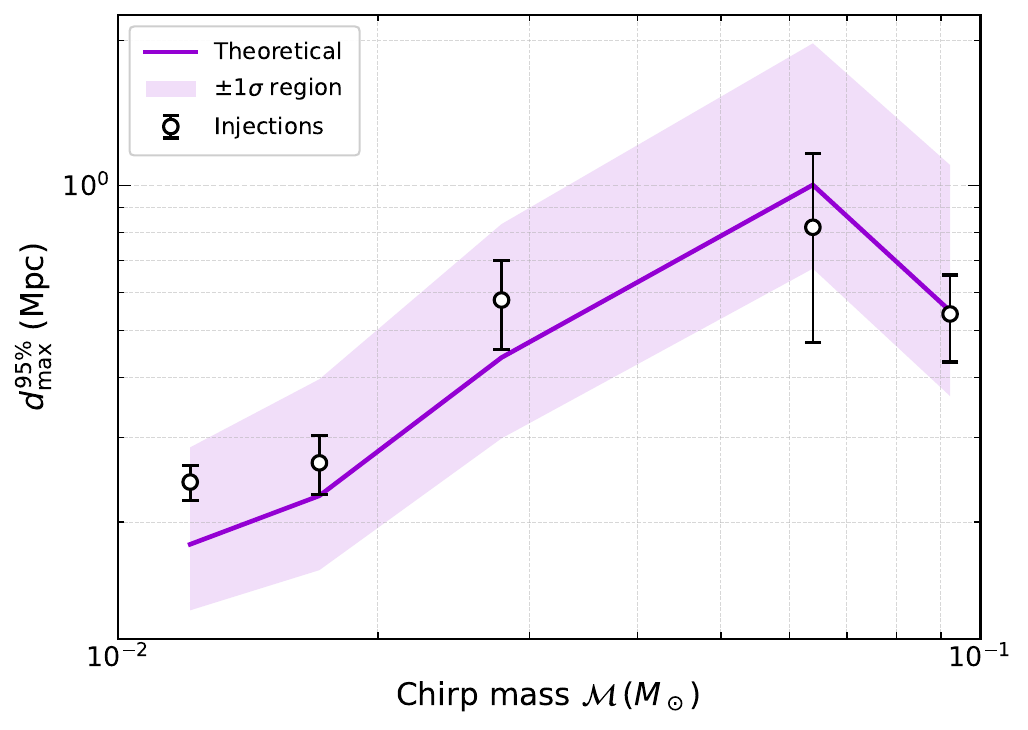}
    \caption{
    Comparison between the measured distance reach of the search and the analytic estimate. The shaded region corresponds to the range of source-specific correction factors applied to the theoretical $\dmax$ to account for the sensitivity expected for individual sources, rather than the population-averaged sensitivity assumed in the analytic expression \cref{eqn:dmax} \cite{KAGRA:2022dwb}. Our data-driven analytic estimate uses the measured \psd from the times at which the signals were injected.
    }
    \label{fig:supp_compare_nonuni_theory}
\end{figure}

\cref{fig:supp_compare_nonuni_theory} compares the analytic
distance reach with those at the 95\% detection probability level obtained through injections using O4a data.
The theoretical values of $\dmax$ correspond to correcting \cref{eqn:dmax} for the source-specific parameters used in each injection \cite{KAGRA:2022dwb}. Since \cref{eqn:dmax} is derived for population-averaged source parameters (in $\cos\iota$, polarization angle, and sky position), we divide it by the mean ratio between the source-specific amplitude and the population-averaged amplitude. The error bars on the theoretical values represent the corresponding $\pm1\sigma$ variation for the 50 injections performed at each distance here. See the appendices of \cite{LIGOScientificCollaborationtheVirgoCollaboration:2025cwh} for further details on this source-specific correction factor. Future studies will improve the \psd estimation in this region.

 In general, we see good agreement between the empirical and theoretical results. We note, however, that for the highest chirp mass considered in the \(\TFFT=0.5\) s configuration, both the empirical and theoretical sensitivities decrease because of difficulties in estimating the power spectral density at such coarse frequency resolution. We experimented with other ways to estimate the \psd, e.g. taking running medians of a collection of adjacent FFTs, but we could not arrive at a way to enhance the sensitivity of this configuration.

\begin{figure}[ht!]
    \centering
    \includegraphics[width=\columnwidth]
    {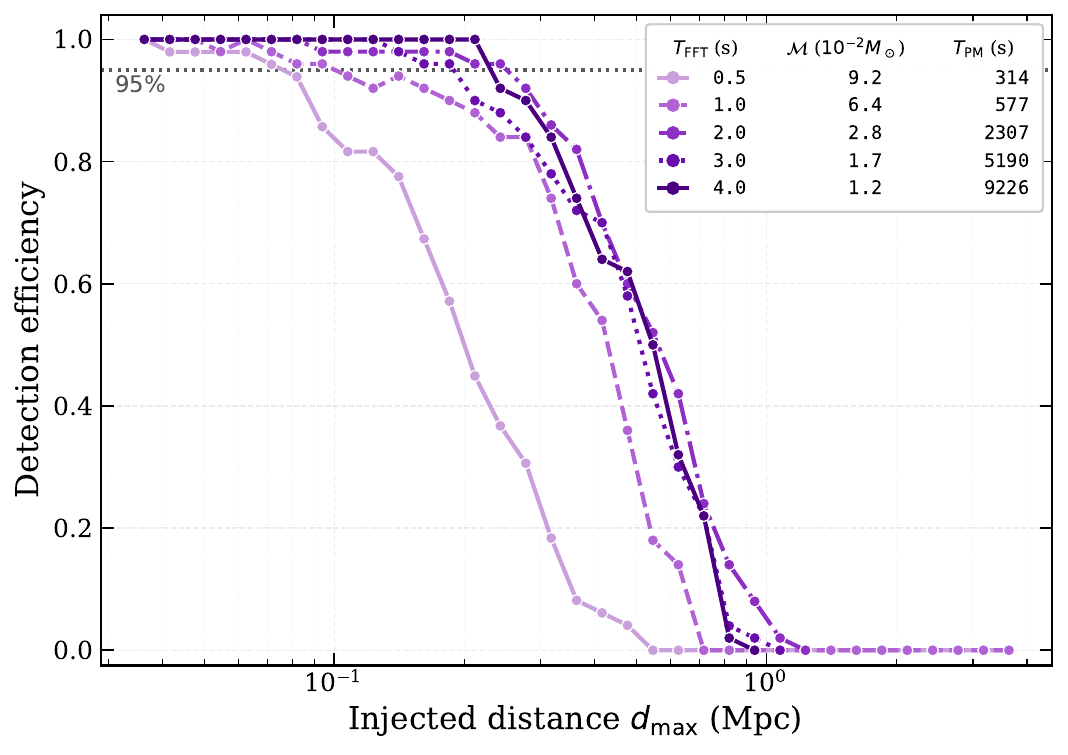}
    \caption{
    Detection-efficiency curves for each search
    configuration. Fifty injections per distance were used here.
    }
    \label{fig:supp_eff_curves}
\end{figure}

 In \cref{fig:supp_eff_curves}, we also provide efficiency curves derived from the O4a injection study that show how the probability of detection scales with distance reach. They tend to follow a sigmoid shape; however, for the $\TFFT=0.5$ s configuration, we see a loss in sensitivity due to the coarse frequency resolution used in the power-spectral-density estimation.
% ============================================================
\section{Injection campaign and sensitive spacetime volume}
\label{app:VT}
% ============================================================

For the rate-density calculation, we simulate inspiraling binaries 
uniformly distributed in spatial volume up to a chirp-mass-dependent maximum
distance $\dmax$. The choice of $\dmax$ balances the impact of truncating the injections at distances too small (downgrading the sensitivity) and the fixed computational power required to perform an extensive injection campaign. If $\dmax$ is too small, nonzero detection efficiency beyond
$\dmax$ is omitted, leading to an underestimate of $\avgVT$.
In contrast, an unnecessarily large $\dmax$ places more injections at distances for which the efficiencies are zero, thus reducing the number of samples of the transition of the efficiency curve for fixed $N_{\rm inj}$.

An injection is counted as
detected when it passes the same $\overline{CR}=5$ sensitivity floor
used for the search follow-up, and when it is in coincidence in both detectors (i.e. within three bins in the $x_0/k$ parameter space).

For a set of $N_{\rm inj}=10^4$ injections at 17 chirp masses, of which $N_{\rm det}$ are
recovered, the detection probability is
\begin{equation}
p_{\rm det}=\frac{N_{\rm det}}{N_{\rm inj}},
\end{equation}
and the sensitive spacetime volume is
\begin{equation}
\avgVT
=
V_{\rm inj}\Tobs p_{\rm det},
\qquad
V_{\rm inj}=\frac{4}{3}\pi \dmax^3.
\label{eq:supp_VT}
\end{equation}
The corresponding 90\% confidence upper limit for zero detected
astrophysical events is \cite{LIGOScientificCollaborationtheVirgoCollaboration:2025cwh}
\begin{equation}
\mathcal{R}_{90\%}
=
\frac{2.303}{\avgVT}.
\label{eq:supp_R90}
\end{equation}

\begin{figure}[ht!]
    \centering
    \includegraphics[width=\columnwidth]
    {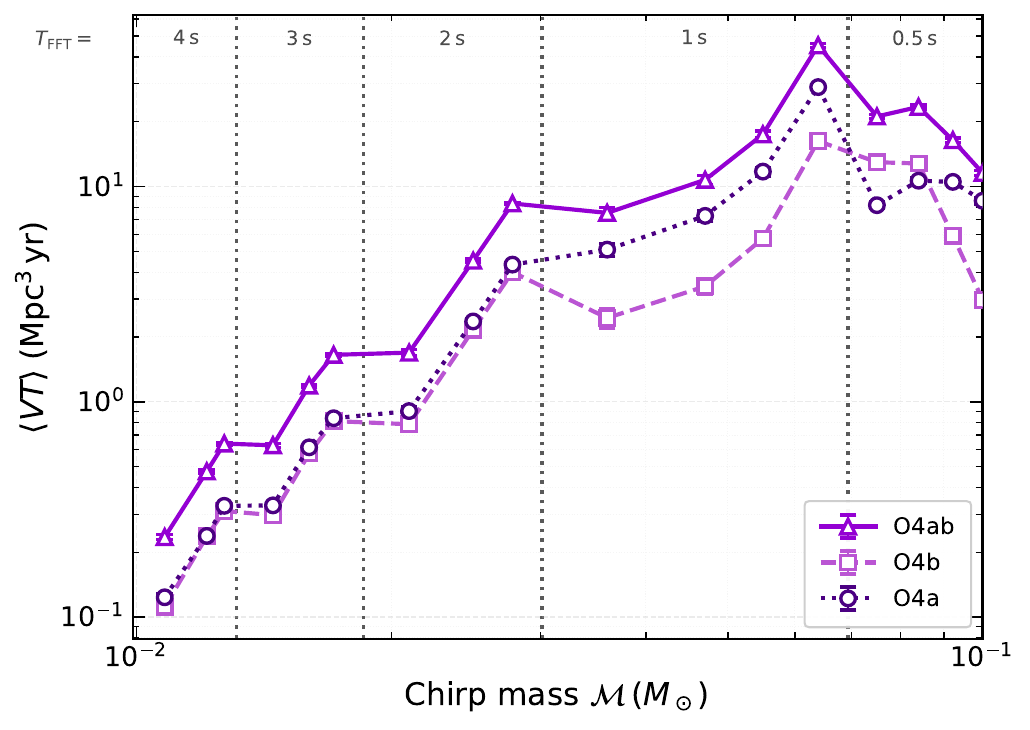}
    \caption{
    Sensitive spacetime volume $\avgVT$ as a function of chirp mass
    for O4a, O4b, and their combination derived through injections.
    }
    \label{fig:supp_VT_O4ab}
\end{figure}

We show in \cref{fig:supp_VT_O4ab} the spacetime volumes that result from the injection campaigns in O4a and O4b, along with their sum that is used to compute $\mathcal{R}_{90\%}$ shown in \cref{fig:R90_high_chirp_mass}.

% ============================================================
\section{Conversion of cosmological merger rates to Galactic rates}
\label{app:cosmo-rates}
% ============================================================

The merger-rate model in \cref{eq:cosmomerg} describes a cosmological
rate density. In contrast to standard \cbc searches, the sensitivity
of our search is restricted to galactic distances.
The cosmological merger rate must therefore be converted to a local
Galactic rate by accounting for the Galactic \dmh density profile.

If the \dmh density in the search volume were approximated by
its value at the Sun's position,
$\rho_{\rm DM}\simeq10^{16}\,\msun\,{\rm Mpc}^{-3}$
\cite{Weber:2009pt}, the enhancement relative to the
cosmological merger-rate density would be  \cite{Miller:2020kmv}
\begin{equation}
    \mathcal{R}
    =
    3.3\times10^{5}\,
    \mathcal{R}_{\rm prim}^{\rm cos}.
\end{equation}
However, the source distances probed by the present search can be
comparable to the distance between the Sun and the Galactic Center.
The assumption of a spatially uniform local \dmh density is therefore
not appropriate. We instead introduce a distance-dependent correction factor $F(d)$ that accounts for the integrated Galactic \dmh density profile over the search volume centered on the Solar position, taken to be $8.2$~kpc from the Galactic Center, and additionally includes the contribution from the M31 dark-matter halo. We model the latter with an NFW profile using the halo parameters inferred from the M31 rotation curve in ~\cite{10.1093/mnras/stae025}, which finds a virial mass $M_{\rm 31}=1.14^{+0.51}_{-0.35}\times10^{12}\msun$, virial radius $r_{\rm vir}=220\pm25$~kpc, and concentration $\log_{10}c=0.94^{+0.25}_{-0.35}$.

Including this correction, we rewrite \cref{eq:cosmomerg} for
equal-mass binaries as
\begin{align}
\mathcal{R}
=&\,
1.04\times10^{3}\,
\mathrm{Mpc}^{-3}\,\mathrm{yr}^{-1}
F(d)
\left(\frac{\mpbh}{\msun}\right)^{-32/37}
\ftilde^{53/37},
\label{eqn:rate-equal}
\end{align}
where $\mpbh=2^{1/5}\Mc$. 

\begin{figure}[ht!]
    \centering
    \includegraphics[width=\columnwidth]
    {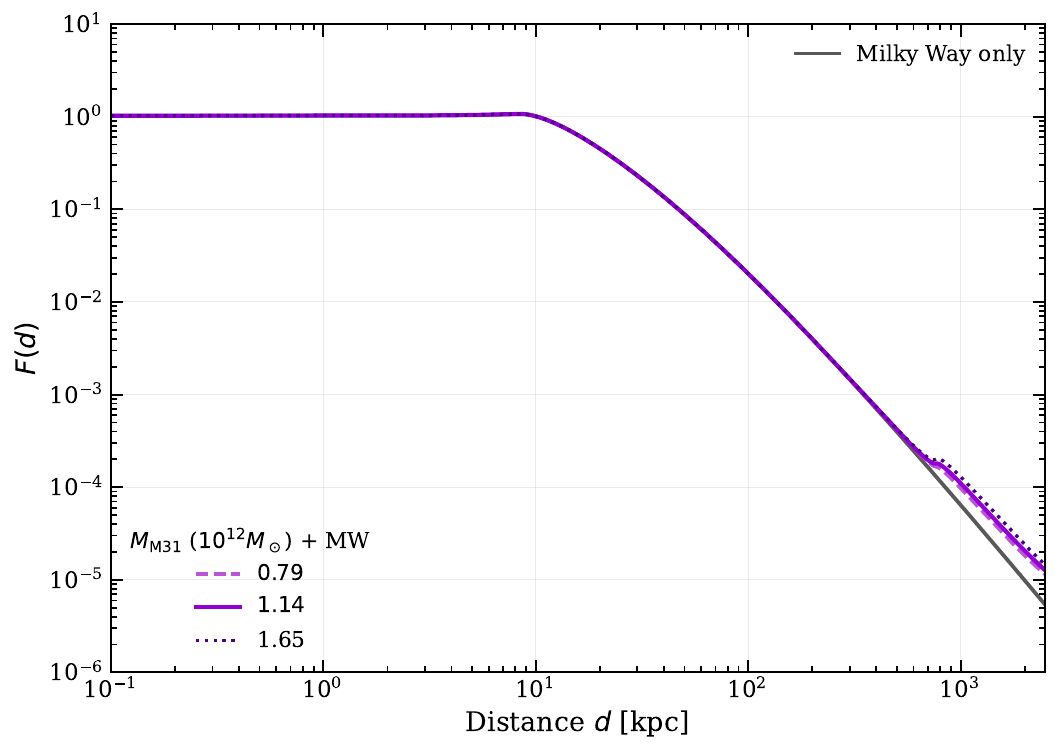}
    \caption{
    Function $F(d)$ that accounts for the fall-off in \dmh density for sources that are located at distances $d$ from us, while including the enhancements from the Milky Way and M31. Colored curves correspond to the contribution to $F(d)$ from M31 by taking the minimum, median and maximum masses from \cite{10.1093/mnras/stae025}, added to the Milky Way (MW) enhancement. We show the curve only considering the MW contribution for comparison, used before in \cite{LIGOScientificCollaborationtheVirgoCollaboration:2025cwh}.
    }
    \label{fig:fofd}
\end{figure}

For asymmetric binaries with $m_2<m_1$,
the corresponding rate is
\begin{align}
\mathcal{R}
&=\,
5.28\times10^{2}\,
\mathrm{Mpc}^{-3}\,\mathrm{yr}^{-1}
F(d) \nn \\ &\times
\left(\frac{m_1}{\msun}\right)^{-32/37}
\left(\frac{m_2}{m_1}\right)^{-34/37}
\ftilde^{53/37}.
\label{eqn:rate_asymm}
\end{align}
The function $F(d)$ is shown in \cref{fig:fofd}. It peaks by a few percent at the galactic center, then decreases with distance from the Galactic Center. When  the function approaches the vicinity of M31, it also peaks slightly, and then falls off again. We define an effective sensitive distance $d_{\rm eff}$ through
\[
\frac{4\pi}{3}d_{\rm eff}^3
=
\frac{\langle VT\rangle}{T_{\rm obs}},
\]
and evaluate the Galactic-density correction $F(d)$ at
$d=d_{\rm eff}$.

% ============================================================
\section{Measured critical-ratio background}
\label{app:CR-background}
% ============================================================

\begin{figure*}[ht]
    \centering
    \includegraphics[width=\textwidth]{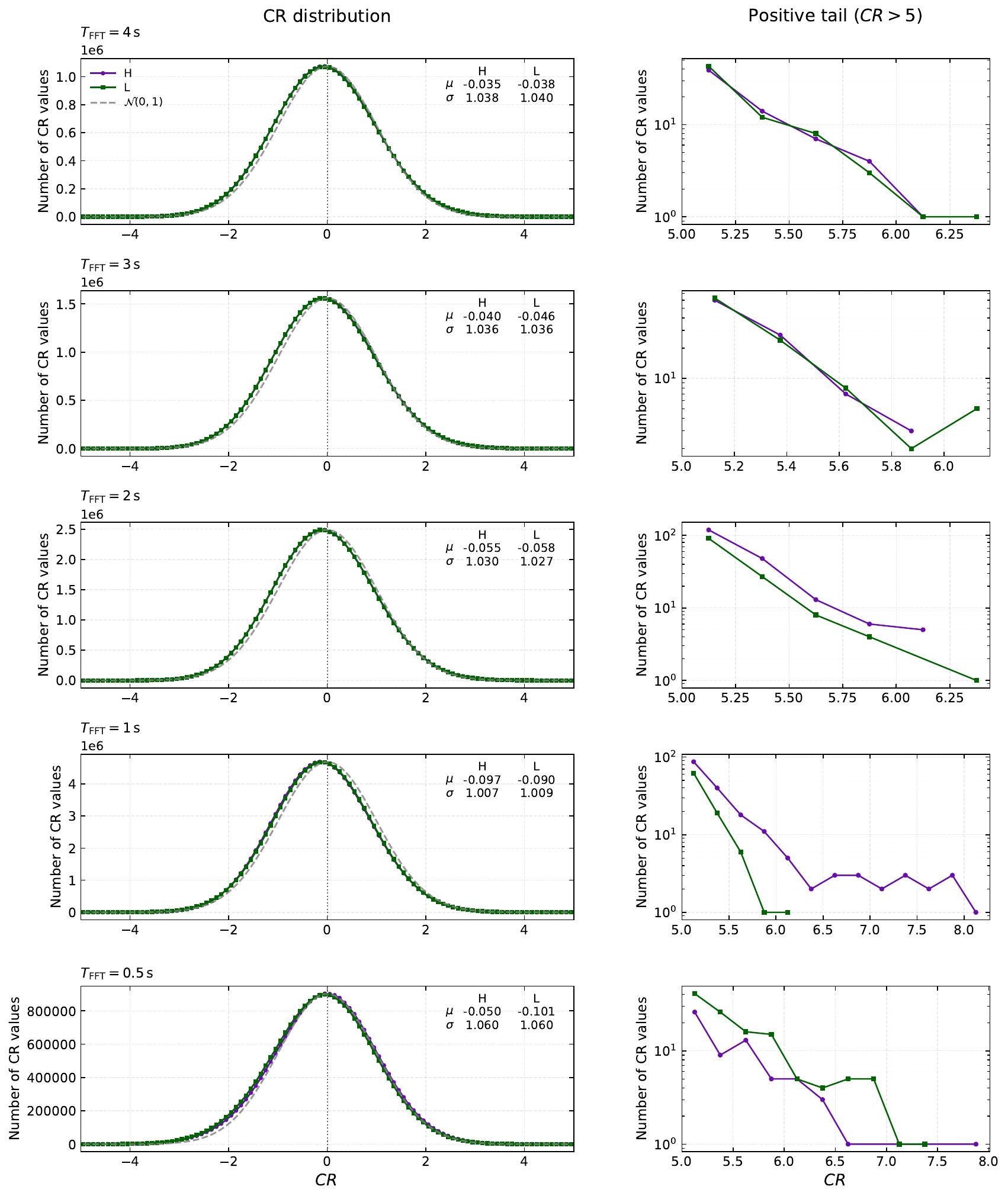}
    \caption{
    Critical-ratio ($CR$) distributions for the five search configurations
    used in the O4a analysis.
    Each row corresponds to a different FFT duration $T_{\rm FFT}$.
    The left panels show the full $CR$ distributions for the H1 and L1 detectors, together with the expected
    standard-normal distribution $\mathcal{N}(0,1)$.
    The measured mean $\mu$ and standard deviation $\sigma$ of each
    distribution are reported in the corresponding panel.
    The right panels show the positive tails with $CR>5$.
    }
    \label{fig:O4a_CR_distributions}
\end{figure*}

\begin{figure*}[ht]
    \centering
    \includegraphics[width=\textwidth]{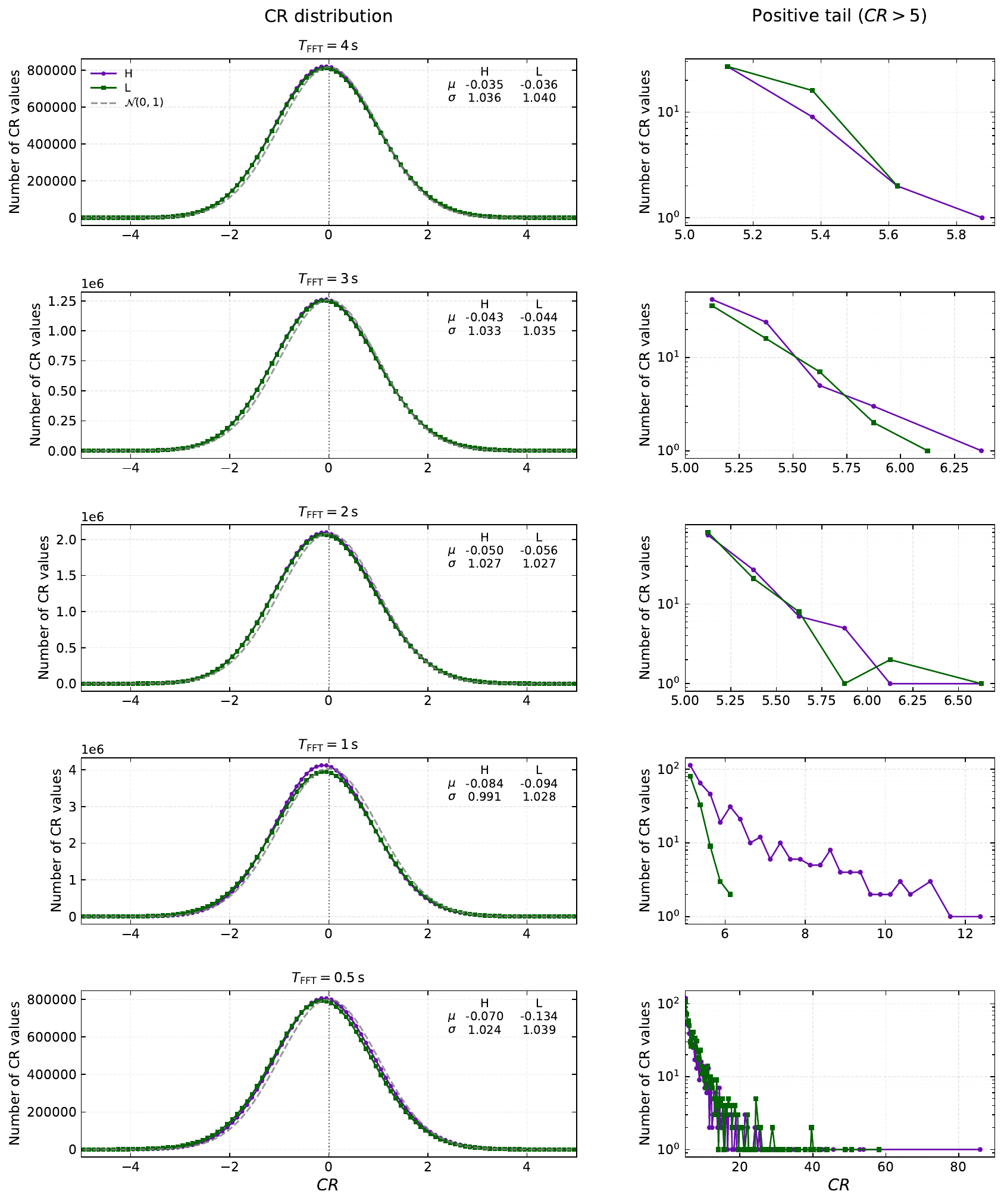}
    \caption{
    Same as \cref{fig:O4a_CR_distributions}, but for O4b.
    }
    \label{fig:O4b_CR_distributions}
\end{figure*}

The \BinGFH critical ratio is constructed to have an approximately
standard-normal distribution in Gaussian noise. In real detector data, we find that a normal distribution is a good fit to the $CR$s obtained from O4ab data for each of the configurations. \cref{fig:O4a_CR_distributions,fig:O4b_CR_distributions} show the measured distributions for
the five search configurations in both O4a and O4b. We see generally good agreement with the standard normal distribution, but their positive tails
contain non-Gaussian disturbances. These non-Gaussian tails, coupled with generally good agreement for the bulk $CR$ distributions, motivate
following up candidates below the nominal trials-corrected threshold
down to $\overline{CR}=5$. Moreover, we note a small negative bias in the means of each configuration away from zero. A bias of $\sim 1\%$ is seen from Gaussian noise simulations; therefore, the majority of the deviation from zero -- up to an order of magnitude larger than the $\sim1\%$ Gaussian-noise bias -- is characteristic of the detector noise. However, this aggregate offset is independent of the real analysis because the per-pixel calibration of individual $CR$ values is computed based on the measured $\mu,\sigma$ shown in \cref{eq:cr,eqn:mu,eqn:sig}.

% ============================================================

% ============================================================

\end{document}